# Water bath calorimetric study of excess heat generation in "resonant transfer" plasmas

**Jonathan Phillips,**[a]

University of New Mexico, Dept. of Chemical and Nuclear Engineering, 203 Farris Engineering, Albuquerque, NM 87131

**Randell L. Mills, and Xuemin Chen**,

BlackLight Power, Inc., 493 Old Trenton Road, Cranbury, New Jersey 08512

Water bath calorimetry was used to demonstrate one more peculiar phenomenon associated with a certain class of mixed gas plasmas termed resonant transfer, or RT plasmas. Specifically, He/$H_2$ (10%) (500 mTorr), Ar/$H_2$ (10%) (500 mTorr), and $H_2O$(g) (500 and 200 mTorr) plasmas generated with an Evenson microwave cavity consistently yielded on the order of 50% more heat than non RT plasma (controls) such as He, Kr, Kr/$H_2$ (10%), under identical conditions of gas flow, pressure, and microwave operating conditions. The excess power density of RT plasmas was of the order 10 W · $cm^{-3}$. In earlier studies with these same RT plasmas it was demonstrated that other unusual features were present including dramatic broadening of the hydrogen Balmer series lines, unique vacuum ultraviolet (VUV) lines, and in the case of water plasmas, population inversion of the hydrogen excited states. Both the current results and the earlier results are completely consistent with the existence of a hitherto unknown exothermic

---

[a] To whom correspondence should be addressed. Phone: (505) 665-2682; Fax: (505) 665-5548; E-mail: jphillips@lanl.gov

chemical reaction, such as that predicted by Mills, occurring in RT plasmas.

# I. INTRODUCTION

Water bath calorimetry studies were performed on certain **novel** gas mixture plasmas containing hydrogen such as $He/H_2(10\%)$, $Ar/H_2(10\%)$, and $H_2O(g)$ called resonant transfer or RT plasmas, and the heat produced was compared to non RT plasmas such as $He$, $Kr$, and $Kr/H_2(10\%)$. The objective was to provide yet another test of a **new** theory of quantum physics [1-17]. Earlier experiments designed to test key aspects of this theory have produced spectroscopic evidence consistent with predictions. The present study, which compares the heat produced by RT and control plasmas run under identical plasma operating conditions, yielded further results consistent with an extraordinary 'chemical' process taking place in RT plasmas. To the present authors, there does not appear to be a 'conventional' explanation of the results based on currently accepted scientific theories.

A key aspect of the **new** theory is the hypothesis that bound electrons are stable (non-radiative) physical particles of specific geometry that obey all of Maxwell's Laws, the requirements of relativity and all the laws of mechanics. For hydrogen atoms, the electrons are "orbitspheres" that surround the nucleus, in quantized states with the same quantized energy levels that satisfy the Schroedinger equation for hydrogen, with significant exceptions. Specifically, the exception are quantum states of hydrogen in which the principal quantum number is a fraction ($n = \frac{1}{2}, \frac{1}{3}, \frac{1}{p}, \ldots \frac{1}{137}$; p = integer replaces the well known parameter n = integer in the Rydberg equation for hydrogen excited states). It is understood that states with the same energies predicted by the **new** theory can also be obtained from the Schroedinger wave equation, using fractional values for n (the principle quantum number), but such 'fractional states' are dismissed as being non-physical, because hitherto there has been no experimental evidence supporting their existence. In contrast, the Mills prediction is not only that there are stable states with these energies, but that it is possible to create conditions in a laboratory plasma capable of effecting transitions to these 'fractional' states. In particular, electrons in the H-atom will fall into these states if certain species are present which singly or multiply ionize at integer multiples of the potential energy of atomic hydrogen, $E_h = 27.2 \, \text{eV}$ where $E_h$ is one Hartree. Specific species (e.g. $He^+$, $Ar^+$, $O_2$, and $2O$) identifiable on the basis of their known electron energy levels are required to be present in plasmas with atomic hydrogen to catalyze the process. Indeed, species such as atoms or ions of $Kr$ or $Xe$ as well as $N_2$ or $CO_2$ do not fulfill the catalyst criterion—a chemical or physical process with an enthalpy change equal to an integer multiple of $E_h$ that is sufficiently reactive with atomic hydrogen under reaction conditions.

Spontaneous emission of photons to create these fractional states is forbidden in the Mill's quantum theory. Catalyst species must be present for the transitions to be observed since the transition requires a nonradiative resonant energy transfer to the catalyst. Moreover, the energy released by transitions to the fractional n=1/p would be very large compared to even the most energetic chemical reactions (e.g. tens of eV vs. <2 eV). The smallest energy change (i.e., $n = 1 \rightarrow n = 1/2$) releases 40.8 eV. Clearly such large energy quanta released in a plasma are capable of producing dramatic effects.

The first of three earlier tests which produced results consistent with the model demonstrated that extreme (>20-200 eV) hydrogen Balmer series line broadening, clearly due to the Doppler effect, is found only in those mixed gas plasmas containing species predicted to have catalytic properties. (It must be noted that the most dramatic evidence was always found with microwave plasmas generated with an Evenson cavity.) Line broadening was not found for any other (non-hydrogen) species in these plasmas. Mills *et al*. [5-6, 14-17] showed that hydrogen line broadening was found in $He/H_2$ (10%), $Ar/H_2$ (10%), and water vapor plasmas as well as certain hydrogen plasmas with gaseous inorganic atoms or ions [2, 4, 9, 13, 16]. Superficially similar plasmas, such as $Kr/H_2$, $Xe/H_2$, and $N_2/H_2$, $CO_2/H_2$, etc. which do not contain species with the catalytic requirements of the Mills model, did not show this broadening. Several other groups also found extreme Doppler line broadening of hydrogen Balmer lines of a similar magnitude in $Ar/H_2$ plasmas [18-24]. These groups proffered explanations, all requiring acceleration of ions toward a cathode, and all peculiar to $Ar/H_2$ plasmas that do not appear to explain the similar broadening found in $He/H_2$, $H_2O(g)$, etc. plasmas. Indeed, no group, other than Mills *et al*., put forth the prediction that the line broadening would be found in any plasma other than the $Ar/H_2$ plasma. Moreover, in recent work it was demonstrated that line broadening can be found only for the Balmer series line in low pressure (<0.5 torr) He/H2 plasmas[25], pure water plasmas[26], and Ar/H2 plasmas [27]. In Xe/H2 plasmas no broadening was found. These results suggest that the broadening is not related to field acceleration, nor the presence of argon.

A second finding consistent with the Mills model is that the unique line broadening is not produced by the fields present in a plasma. Earlier groups specifically postulated that the extreme Doppler hydrogen line broadening found in mixed gas plasmas is due to $H_2^+$ acceleration by the cathode, and subsequent neutralization, dissociation, and emission. Yet, Mills *et al.* observed the line broadening in (electrodeless) microwave plasmas [15-17]. Moreover, in a pure water plasma, the line broadening (undiminished in magnitude) persists at a distance of more than 5 cm from the microwave coupler [28]. Given that ion acceleration is absolutely minimal in microwave systems, and that the strong fields are confined almost entirely to the region of the coupler, these findings indicate that the process that creates the broadened lines is a local one that can take place in low field regions. This conclusion is further strengthened by the finding in water plasmas that population inversion of atomic hydrogen excited states ('pumping') is also found at distances greater than 5 cm from the microwave coupler. Given that the excited hydrogen states producing the Balmer lines have lifetimes of the order of $10^{-7}$ s (which translates to less than 1 mm of travel) it is clear that the excited state populations observed spectroscopically are created at the point of observation. Again, these findings are consistent with a **novel** 'chemical' process involving hydrogen and a catalyst. To the authors of this study it does not appear that there is a conventional explanation sufficient to satisfy all of these observations simultaneously.

A third finding consistent with the **novel** model, one which is particularly compelling, is that the VUV spectral lines *predicted* by the Mills model are present at the predicted wavelengths in RT plasmas and not in any similar control plasma [5-7, 10-12]. There appears to be no conventional explanation for the **novel** VUV lines, which occur only in mixtures containing hydrogen and one of the Mill's predicted catalyst gases.

The present study was designed to test a fourth prediction of the model, namely the possibility of the transitions of electrons to deep lying fractional quantum states producing measurable amounts of thermal energy. Significant heat is released in RT plasmas by the process of electron transitions to fractional n-states effected by a catalyst which resonantly accepts $m \cdot 27.2\,\text{eV}$ wherein m is an integer. A simple formula provides the energy of transition to each of these states [7]:

$$H\left[\frac{a_H}{p}\right] \rightarrow H\left[\frac{a_H}{(p+m)}\right] + [(p+m)^2 - p^2] \times 13.6 \text{ eV} \qquad (1)$$

where *p* and *m* are integers and $n_i = 1/(p+m)$ designate the initial and final energy states corresponding to the principal quantum number n in the Rydberg formula.

Clearly even the lowest energy transition of the electron of atomic hydrogen to a fraction Rydberg state releases significant energy that can be observed as heat. Specifically, only the Mills model predicts that the energy output of RT plasmas should be higher than that of controls for identical power inputs due to the catalytic process. Thus, the present experiment was designed to compare/contrast the heat output to a water bath by control and RT plasmas operated under identical conditions to test the so-called *Mills hypothesis of heat production*.

This last hypothesis, if true, provides the basis for interest in this subject for the applied physics community. If heat can be released via this process in a reproducible and consistent fashion, then transitioning hydrogen to fractional states represents a truly **novel** source of chemical energy. In fact, on a per hydrogen atom basis, this process is predicted to produce tens to hundreds of times more energy than any combustion process. And, as discussed below, the present results are consistent with the following interpretation: energy was produced at a steady rate of about 10 W/cm$^3$, via transitions of hydrogen to fractional quantum states, over a small volume of the plasma, for all the RT plasmas studied.

## II. EXPERIMENTAL

A unique apparatus was designed for this experiment. The design goals of the equipment were first, to allow high precision, reliable, absolute measurements of the amount of heat output from Evenson-cavity-generated microwave plasmas, and second to quantify the microwave power to the plasmas. Meeting these goals allows a proper test of the Mills heat production hypothesis.

The first goal was achieved by using a water bath calorimeter (Figures 1-2) designed such that the entire discharge tube was submerged. A water bath instrument was judged superior to alternative calorimeter designs as all power generated by the plasma is absorbed by the water, there is no electrical signal that might somehow be corrupted, no 'partial' signal that must be integrated, etc.

The water bath calorimeter described previously [29] was calibrated using a precision electrical resistor circuit (Watlow 125A65A2X, powered with a Xantrex DC power supply, 0-1200 +/-.01 W) permanently installed within the water tight cavity surrounding the Evenson cavity coupler (Figures 1-2). Repeated calibration experiments were conducted at 25-100 W inputs which essentially measured the system heat capacity. The heat capacity was found, as expected, to be independent of the power input, and the precision of the measurement was found to be +/-0.5%. It should be noted that the temperature rise of the 45 liter, distilled-water bath (always linear, see Figure 3) was measured both with a mercury thermometer with a resolution of 0.05 K, and a linear-response thermistor probe (Omega OL-703) with a precision of +/-0.001 K. Both measures provided absolute calibration (heat capacity, $C_p$) constants (J/°K) within 2% of that calculated on the basis of the mass of water in the bath and the amount of glass and metal in the Evenson cavity cage and quartz discharge tube. As the latter values are estimates, the calibrated values were used. Subsequent heating rates are reported on the basis of computer-collected thermistor readings, although in most cases, mercury thermometer readings were made at regular intervals as an added check.

The energy balance for a system consisting of the contents of the water bath calorimeter is

$$\dot{H} = \dot{M}(\hat{H}_{in} - \hat{H}_{out}) + \dot{Q}_{plasma} + \dot{Q}_{power\,cable} + \dot{Q}_{stirrer} + \dot{Q}_{heat\,exchange} \qquad (2)$$

where H's are enthalpy values (inlet and outlet gases as indicated by the subscripts in and out, respectively, and the hat designates per mole), $\dot{M}$ is the molar flow rate, and the $\dot{Q}$'s are heat flow rates. It is clear from Eq. (2) that a correction must be considered both for the gas flow term (first term, right side), '$\dot{Q}_{power\,cable}$' which represents the input of the short section (approx. 3 cm length) of the coaxial cable housing that passes through the water bath as it brings power to the microwave power coupler, for the work of the stirrer, and for the heat exchange between the insulated water bath and it surroundings.

The values of '$\dot{Q}_{power\,cable}$' and the heat carried out with the gas were small, as determined by appropriate temperature readings. Thermocouples were employed to measure the temperature of the input and output gas, as well as the temperature of the coaxial cable housing just outside the water bath. Given that the temperature of the cable housing was never more than 3 K higher than the water bath, $\dot{Q}_{power\,cable}$ can readily be shown to be negligible. The gas temperature change between input to the plasma and output from the water bath was never more than 1 K. Heat transfer from the quartz tube containing the flowing gas to the water in the bath was clearly very efficient. Given the maximum flow rate was 10 sccm, this requires a maximum correction of less than $10^{-5}$ W, a trivial correction. The stirrer and heat exchange terms were found to be the most significant correction, but its value was readily determined by measuring the temperature rise with only the stirrer operating. This correction can be accurately calculated from the slope of the pre- and post-heating periods and was found to be constant and of the order of 1 W for all experiments. Once these relatively trivial corrections are made the 'effective' energy balance becomes:

$$\dot{H} = \dot{Q}_{plasma} \qquad (3)$$

The calibration procedure resulted in a linear change in temperature for constant power inputs. This is expected given the nearly constant heat capacity of water over small changes in temperature (<3 K in all cases). Thus, changes in enthalpy can be readily equated with change in temperature of the bath. In short:

$$\dot{H} = C_p \dot{T} = \dot{Q}_{plasma} \qquad (4)$$

Thus, one must only multiply the calibration constant by the rate of change of bath temperature to obtain the plasma's heating power of the water bath. In the event that the change in temperature is nearly linear with time, as it was in all cases in this study, the rate (W) of heat input from the plasma to the bath can be readily determined, and compared with the microwave input power.

The microwave plasmas were generated, in a 1.27 cm OD U-shaped quartz tube about 20 cm in length (Figures 1-2), using an Opthos model MPG-4M generator (2.45 GHz, E-mode) as described previously [26]. The glowing plasma volume was about 3 cm$^3$ in the metal encased U-tube, about one-half that observed in 'open' straight tubes where the cavity was not surrounded by a metal housing. Gas pressure (other than for some water vapor and Kr studies) was maintained at 500 mTorr, measured with a 0-10 Torr MKS absolute pressure Baratron, using a molecular drag pump. Flow rates, other than for water vapor, were maintained at 10 sccm with 0-20 sccm mass flow controllers (MKS). For the water plasmas, weight loss measurements showed the flow rates to be about 1 sccm.

The second goal of quantifying the microwave input power to the plasmas was achieved by maintaining forward and reflected powers, as measured with power diodes, identical in all cases. That is, for both the RT and control plasmas, the power supply and Evenson cavity tuning were adjusted such that the forward and reflected powers indicated by the power reading diodes (Agilent) of the Opthos generator were identical.

In general, the ultimate basis of any microwave power measurement is a direct absolute calorimetric determination on an instrument which then serves as a primary standard for secondary measurements. In our experiments, that absolute standard was the water bath calorimeter at fixed forward and reflected power diode settings. For example, with the assumption that krypton plasma (non RT plasma) does not produce excess heat, the power dissipated in the load is absolutely known at particular fixed forward and reflected power diode readings from the calorimetric measurement. Thus, if the diode values are matched identically for any other plasma load, the power dissipated in that load must be identical since the system and the measured power flows are identical. This was tested by running 40 control non RT plasmas of various pure gases and gas mixtures (e.g. $He$, $Ar$, $Kr$, $Kr/H_2$, $Xe$, $Xe/H_2$, $N_2$, $N_2/H_2$, $CO_2$, $CO_2/H_2$) with different pressures and mixture ratios. The resulting common water bath reading served as the calibration of the input power to RT plasmas ($He/H_2(5\%)$, $He/H_2(10\%)$, $Ar/H_2(3\%)$, $Ar/H_2(10\%)$, and $H_2O(g)$). That is, since presumably the electron density, electron temperature, ionization fraction, hydrogen atom density and energy, and gas temperature are different for all plasmas, if these factors impact the power diode readings, then each different control plasma, even for constant diode readings, would produce a different water bath signal. Thus, the nearly identical heat signals measured for the many control plasmas demonstrate the reliability of the power diodes, set at constant readings, as a means to create identical net power input from the magnetron to each plasma. It should also be noted that each plasma required unique 'stub tuning' to produce the desired forward and reflected power readings. Thus, it is clear that 'tuning' also does not impact the readings of the power diodes.

Thorough tests were conducted to demonstrate that the magnetron only broadcasts energy at a single frequency in order to demonstrate that the diode methodology for measuring forward power is consistent for all plasmas, both control and RT, studied. That is, if for RT plasmas some form of "feedback' corrupts the output of the magnetron, differently than for control plasmas, then off-resonance frequency energy may enter the plasma for RT plasmas only and not be "registered" by the power diodes. To determine if off-frequency electromagnetic emission existed for any of the plasmas, spectral measurements were conducted for all of the control and all of the noble gas/hydrogen (RT) plasmas over the full frequency range of the spectral analyzer (0-10 GHz) and over the relevant range of forward (60 +/- 20 W) and reflected (up to 20 W) powers of the microwave generator. No signal (e.g. no off spectral resonances) except at 2455 MHz was detected over this range.
Detailed analysis was then conducted within +/- 10 MHz of the resonant frequency, 2455 MHz. In all cases the shape of the resonant peak was the same and no abnormal emissions were detected.

The above analysis was done using a 6-turn Ni wire coil antenna symmetrically wrapped around the outside of the plasma tube, centered on the Evenson cell. The power spectrum of the microwave power driving the plasmas was analyzed using an Advantest R3265 spectrum analyzer. The power spectrum into all plasmas under the identical operating conditions as the calorimetric studies was determined to be a single peak at 2455 MHz with a full-width-half-maximum of 1 MHz. The relative input power at different diode settings for the forward and reflected power was determined by calibrating the peak height relative to a microwave power meter. The power diodes were determined to respond linearly over a broad range of forward and reflected powers (+/- 20 W) around the particular absolutely calibrated diode readings that were matched in each experiment. Thus, slight mismatches (on the order of +/-1 W) were determined to be inconsequential to the results. Given the high quality of the magnetron and associated control equipment this result is not surprising.

## III. RESULTS AND DISCUSSION

The results of this study clearly show that the heat generated by RT plasmas ($He/H_2$, $Ar/H_2$, $H_2O(g)$) was greater than that generated by non RT plasmas ($He$, $Ar$, $Kr$, $Kr/H_2$, $Xe$, $Xe/H_2$, $N_2$, $N_2/H_2$, $CO_2$, $CO_2/H_2$) for the same plasma system operating parameters as shown in Table I for 500 mTorr plasmas and Table II for 200 mTorr plasmas. The nominal values to which all the test plasmas could be tuned was 70 +/-2 W forward, 16 +/-2 W reflected. The calorimetrically measured power dissipated in the control plasmas matched 37.5 +/- 2.5 W in all cases as shown for some representative results given in Tables I and II. The range of readings for the non RT plasmas of +/- 2.5 W probably does not reflect any systematic differences between control plasmas, as even repeated measures of one gas ($Kr$) produced a similar range of variation (see statistical analysis below). The 2.5 W variation probably reflects the true net uncertainty in the experiments. From these results, power input to each RT plasma ($He/H_2$, $Ar/H_2$, $H_2O(g)$) was confidently known as the diode readings identically matched those of the controls. That is, there is no reason to believe that the power input into the RT plasmas is any different than that input to the ten different 'control' plasmas.

The results of the 1Torr RT plasmas are given in Table I. A representative result for the absolute calibration of the input power with a $Kr$ plasma and the higher output power observed with a $He/H_2$(10%) RT plasma is shown in Figure 4. There is remarkable consistency in the findings: 500 mTorr RT plasmas consistently produced between 55 and 60 W in the water bath. Only the $He/H_2$(5%) plasmas were slightly lower (approx. 47 W). With an uncertainty of +/- 2.5 W, it is clear that the consistent difference in output power (20 W, >50%) between RT and control plasmas is far outside the range of either theoretical or measured error ranges (see statistical analysis below).

In companion studies of water plasmas generated with an identical Evenson cavity in a matched quartz tube, it was found that the discharge length for 200 mTorr plasmas is about double that found at 500 mTorr. Thus, we decided to test the heat output from 200 mTorr water plasmas. That is, a longer discharge implies a longer period for 'chemical RT processes' to occur, hence generation of additional heat. As shown in Table II and Figure 5 this increased the average water plasma heat output to more than 60 W repeatedly. In contrast, Kr plasmas run at this pressure produced heat output in the same range as all the 500 mTorr control plasmas.

The net input power of 54 W given by taking the difference of the diode readings of the forward and reflected powers shown in Table I is of no particular significance. These diodes were only single-point calibrated at the manufacturer and may not have been accurate. And, even with calibration in a water bath calorimeter, they are only absolute in the absence of a high reflected microwave power. It is clear, however, that leaving the settings constant even in the presence of a high reflected wave as in our case should result in a consistent supply of power to the plasma. The readings of these diodes should then be interpreted as indicating a fixed input condition rather than representing the absolute net power delivered. Indeed water bath calorimetry is an absolute standard and indicates 37.5 +/- 2.5 W input power at the diode settings 70 W forward and 16 W reflected. Clearly, for the wide range of control gases and control gas mixtures studied, the diode readings were not influenced by the plasma gas composition, or even the pressure. That is, for all the control gas combinations, 35 to 40 W was measured as the input power to the water bath; whereas, for the RT plasmas an average of about 50% more thermal power was observed.

Furthermore, given a mass flow rate of 10 sccm, the energy balance corresponding to the excess power of the RT plasmas was very high. Since the total gas flow rate was about $7.5 \times 10^{-6}$ $moles/s$ and the excess power averaged about 20 W, the average 'excess' energy was $3 \times 10^4$ $kJ/mole\ gas$ (or approx. 150 eV/hydrogen atom), compared to the energy of combustion of hydrogen of $-241.8\ kJ/mole\ H_2$. Moreover, the only known exothermic chemistry for the gases and gas mixtures of this study is for some of the controls. The enthalpy for the reaction of hydrogen and nitrogen to form ammonia [30] is

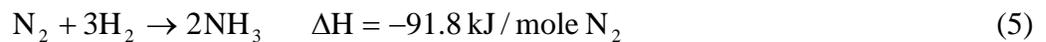

$$N_2 + 3H_2 \rightarrow 2NH_3 \quad \Delta H = -91.8\ kJ/mole\ N_2 \quad (5)$$

The enthalpy for the reaction of carbon dioxide with hydrogen to form methane and water [27] is

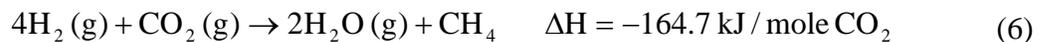

$$4H_2(g) + CO_2(g) \rightarrow 2H_2O(g) + CH_4 \quad \Delta H = -164.7\ kJ/mole\ CO_2 \quad (6)$$

With a flow rate of 10 sccm, the power from the reactions given by Eqs. (5) and (6) is negligible, about 0.1 W. The very large energy 'excess' observed with the RT plasmas comprising a catalyst and atomic hydrogen were consistent with the formation of fractional Rydberg states of atomic hydrogen.

Even though the energy production rate data (either table) for the non-RT plasmas and the RT plasmas does not overlap, a thorough statistical analysis was performed. Standard one-sided t-tests of the null hypothesis, that the means are equal vs. the alternative hypothesis that the RT data is greater than the non-RT data, were carried out. In both cases, the test clearly rejects the null hypothesis that the means are equal. The test assumes equal variances, which may not be true for the Table 1 data because of the two aberrant observations for He/H2 (5%) plasmas.

However, because the data don't even overlap, an alternative Wilcoxon nonparametric test based on the ranks of the observations in the two groups would also clearly reject the null hypothesis. (Probability of obtaining Wilcoxon statistic as large as the calculated value is .003 for Table 1 and 1.1e-6 for Table 2. That is, the probability that the two data sets are actually equivalent is less than one third of one percent.)

For Table 1, the mean of the control plamas (non-RT) power values is 37.9, whereas the mean of the RT plasma power values is 56.6. The t-statistic for the above t-test is 16.8 with 32 degrees of freedom and a p-value of zero (to several decimal places). That is, the probability that a t-statistic at least this large would be obtained if the means were actually equal, is extremely small. The estimated difference in the means is 18.7, with a 95% one-sided lower confidence bound of 16.8.

For Table 2, the mean of the controls (non-RT) power values is 38.8, and the mean of the RT power values is 62.7. The t-statistic for the above t-test is 15.1 with 9 degrees of freedom and a p-value of 5.3e-8. Clearly, the probability that a t-statistic at least this large would be obtained if the means of the two data sets were really equivalent is virtually zero. The estimated difference in the means is, 23.9, with a 95% one-sided lower confidence bound of 21.0.

## IV. CONCLUSIONS

This study was carried out to further test the Mills' hypothesis of heat production in plasmas satisfying the resonant transfer (RT) criteria. *In appropriate RT plasmas hydrogen electrons will fall into fractional quantum states due to catalytic reactions between hydrogen atoms and a catalyst. This will result in the generation of a significant amount of energy.* The results of our study were shown to be consistent with the Mills' hypothesis but it should be fully understood that these results do not establish the correctness of the **new** Mills theory.

Specifically, water bath calorimetry studies showed that low pressure RT microwave-plasmas generated with an Evenson cavity consistently produced on the order of 50% more heat than non RT plasmas run under identical conditions, although the only difference between the RT and non RT plasmas was the identity of the gases. Furthermore, the corresponding excess power density of RT plasmas was high, of the order of $10 \: \text{W} \cdot \text{cm}^{-3}$.

The incentive for carrying out the study, and the design (e.g. use of low pressure Evenson cavity generated plasmas) were based on the surprisingly strong support for the Mills hypotheses provided by several spectroscopic tests. Those very rigorous tests produced results consistent with the Mills model, but having no satisfactory explanation in standard QM theory. In particular, previous spectroscopic investigations demonstrated that low pressure RT plasmas generated with Evenson microwave cavities have many of the spectral features (e.g. specific spectral lines in the VUV region) predicted by the Mills model [5-7, 10-12]. The present work represents the first thorough test of a non-spectral characteristic of RT plasmas, heat production.

The authors can find no satisfactory conventional physics explanation for the experimental results reported herein. That is, it is not clear what special feature of the RT plasmas gases could cause more power to be generated and subsequently cause the water bath to heat up at a rate more than 50% greater than matched controls. The standard model of microwave plasmas allows only for electrons to absorb power from the microwave field. In turn, this absorbed energy is partially converted by various mechanisms into electron excitation, vibration, rotation, and translation of the various atomic and molecular species present. All the energy initially absorbed by the electrons from the microwave field is eventually transferred to the water in the bath. Moreover, the reliability of the power diodes as a means to determine input power is attested to by the remarkable similarity of water bath behavior for more than 10 different control plasmas, each presumably with a distinct EEDF, radical population and even gas temperature profile. Each control plasma required a unique 'tuning' of the two stubs to obtain the desired forward and reflected power readings. That is, the 'reflectivity' of the plasma did not disturb the net input power determination of the power diodes.

There is no known chemistry possible in a $He/H_2$, $Ar$, or water vapor plasma known to the authors that would yield the significant 'excess' power observed. There is no standard physics explanation for the finding that extra heat is associated only with those plasmas that produce dramatic Balmer series line broadening [5-6, 14-17], inversion of the population of excited states [14, 28], and specific spectral lines in the EUV region [5-7, 10-12]. In contrast, these findings are consistent with the Mills model *predictions* that RT plasmas will generate heat, in addition to that absorbed by electrons from the microwave field, through the catalytic reaction with atomic hydrogen that cause electrons in hydrogen atoms in RT plasmas to transition to states with fractional quantum numbers.

It bears repeating that the results of this study do not 'prove' the correctness of the new Mills model of QM. Still, the scientific method requires that a theory that correctly predicts multiple outcomes must be seriously considered. Further testing by the scientific community for 'excess heat' and other phenomena associated with the **new** Mills' theory is clearly warranted. One possible test suggests itself immediately: collect the 'gas' product of the RT plasmas and test it for unusual spectroscopic, NMR and other features.

Table I. The water bath calorimeter response to RT plasmas ($He/H_2$, $Ar/H_2$, $H_2O(g)$) and non RT plasmas ($He$, $Ar$, $Kr$, $Kr/H_2$, $Xe$, $Xe/H_2$, $N_2$, $N_2/H_2$, $CO_2$, $CO_2/H_2$) at 500 mTorr for the same plasma system operating parameters.

| Plasma Gas | Forward Power (W) | Reflected Power (W) | Water Bath Power (W) |
|---|---|---|---|
| Kr | 70 | 16 | 35.3 |
| Kr | 70 | 16 | 37.5 |
| Kr | 70 | 16 | 38.8 |
| Kr | 71 | 17 | 37.1 |
| Kr | 71 | 17 | 37.8 |
| Kr | 72 | 18 | 35.9 |
| Kr | 70 | 14 | 38.7 |
| $Kr/H_2$ (5%) | 70 | 16 | 36.7 |
| $Kr/H_2$ (5%) | 70 | 16 | 37.6 |
| $Kr/H_2$ (20%) | 70 | 16 | 37.8 |
| Xe | 70 | 16 | 35.2 |
| $Xe/H_2$ (20%) | 70 | 16 | 37.2 |
| $CO_2$ | 70 | 16 | 40.5 |
| $CO_2/H_2$ (20%) | 70 | 16 | 39.7 |
| $N_2$ | 71 | 17 | 37.5 |
| $N_2$ | 70 | 16 | 37.9 |
| $N_2$ | 70 | 16 | 40.3 |
| $N_2 / H_2$ (20%) | 70 | 16 | 40.2 |
| Ar | 70 | 16 | 36.9 |
| Ar | 68 | 14 | 38.8 |
| $Ar / H_2$ (3%) | 70 | 16 | 60.5 |
| $Ar / H_2$ (3%) | 70 | 16 | 60.6 |
| $Ar / H_2$ (10%) | 70 | 14 | 60.5 |
| He | 70 | 16 | 38.6 |
| He | 72 | 18 | 37.6 |
| $Ar / H_2$ (5%) | 70 | 16 | 47.4 |
| $Ar / H_2$ (5%) | 70 | 16 | 47.8 |
| $Ar / H_2$ (10%) | 68 | 14 | 60.6 |
| $Ar / H_2$ (10%) | 70 | 16 | 59.2 |
| $Ar / H_2$ (10%) | 70 | 14 | 59.3 |
| $Ar / H_2$ (10%) | 70 | 17 | 55.6 |
| $H_2O(g)$ | 70 | 16 | 52.6 |
| $H_2O(g)$ | 70 | 16 | 58.8 |
| $H_2O(g)$ | 70 | 16 | 56.2 |

Table II. The water bath response to $H_2O(g)$ RT plasmas and Kr non RT plasmas at 200 mTorr for the same plasma system operating parameters.

| Plasma Gas | Forward Power (W) | Reflected Power (W) | Water Bath Power (W) |
|---|---|---|---|
| Kr | 70 | 15 | 39.9 |
| Kr | 70 | 16 | 38.5 |
| Kr | 70 | 16 | 37.1 |
| Kr | 70 | 16 | 39.8 |
| $H_2O(g)$ | 70 | 15 | 68.5 |
| $H_2O(g)$ | 70 | 15 | 62.1 |
| $H_2O(g)$ | 70 | 15 | 62.7 |
| $H_2O(g)$ | 70 | 16 | 60.4 |
| $H_2O(g)$ | 70 | 16 | 59.3 |
| $H_2O(g)$ | 70 | 16 | 62.3 |
| $H_2O(g)$ | 70 | 17 | 63.7 |

Figure Captions

FIG. 1. Schematic of the water bath calorimeter for measuring the power generated by pure gases and gas mixtures. The Evenson cavity and a plasma-containing section of the quartz tube were fitted with a water-tight stainless steel housing, and the housing and cell assembly were suspended by 4 support rods from an acrylic plate which held the cell vertically from the top of a water bath calorimeter.

FIG. 2. Schematic of the water bath calorimeter for measuring power on water vapor plasmas. Only the source of gas differs from Figure 1.

FIG. 3. The thermogram, $T(t)$ response of the cell, with stirring only and with a constant input power to the high precision heater of 50.0 W. The baseline corrected least squares fit of the slope, $\dot{T}(t)$, was $2.622 \times 10^{-4} \,°C/s$, and the heat capacity was determined to be $1.907 \times 10^5 \, J/°C$.

FIG. 4. The $T(t)$ water bath response to a sequence of inputs. First, only the power of stirring was recorded. Next the power input to the water bath from a $Kr$ only plasma with the standard panel meter readings of 70 W forward and 15 W reflected microwave power was recorded. The microwave input power was determined to be 38.7 +/- 1 W. After a second period of simple stirring and gas replacement, a $He/H_2(10\%)$ RT plasma (rapid retuning required) was run at identical microwave input power readings as the control, and the power was determined to be 59.3 +/- 1 W from the $T(t)$ response.

FIG. 5. The $T(t)$ water bath response to two distinct plasmas operated at 200 mTorr (Table II), and the standard forward and reflected power settings are shown. The Kr only plasma produces 37.1 +/- 1 W, within the range observed for all the controls. In contrast, the water only plasma at this pressure produced 68.5 +/- 1 W.

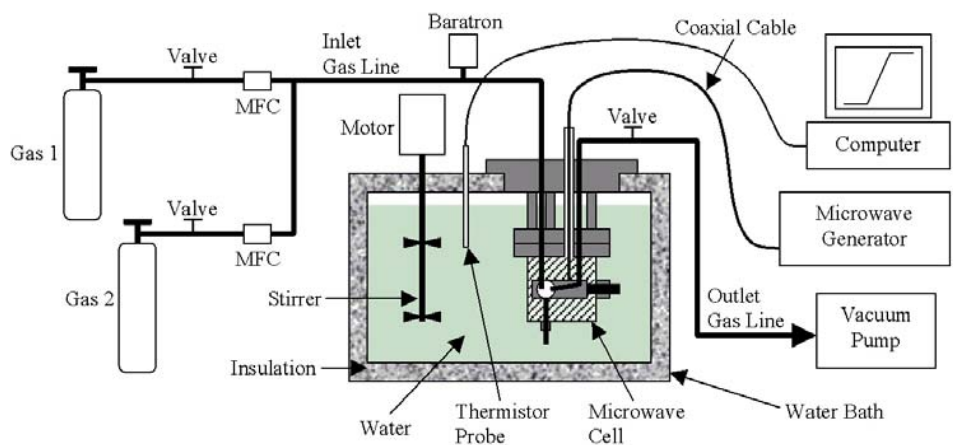

Fig. 1

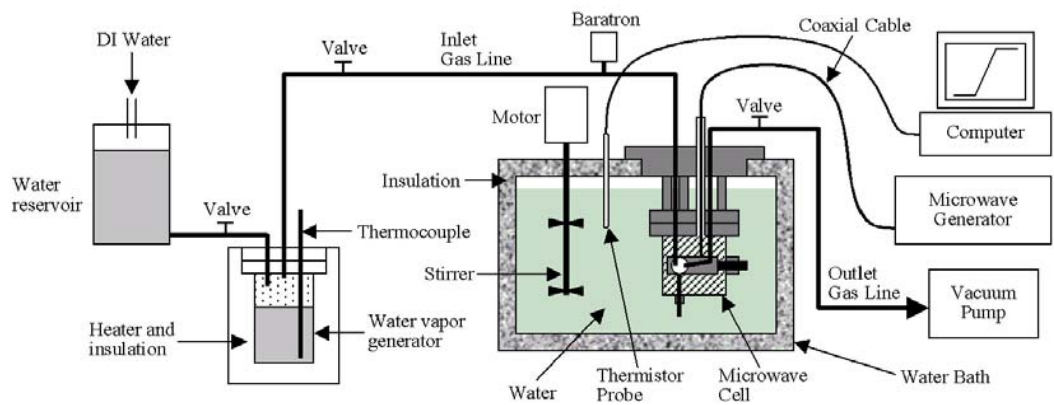

Fig. 2

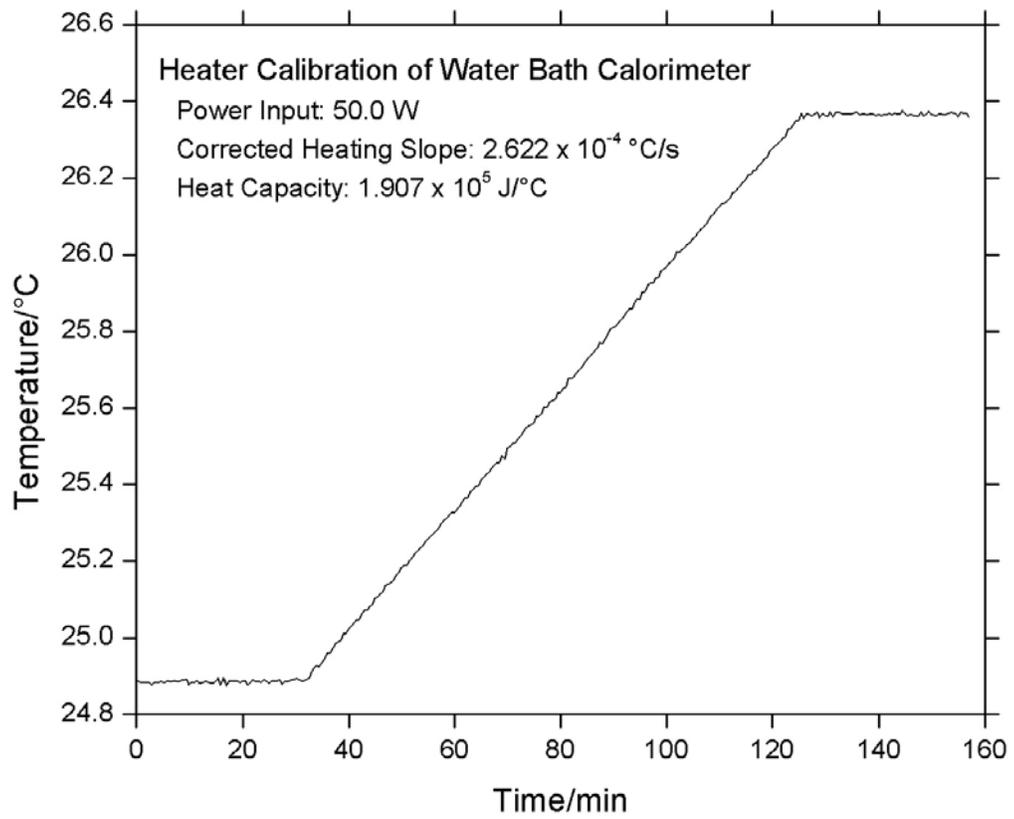

Fig. 3

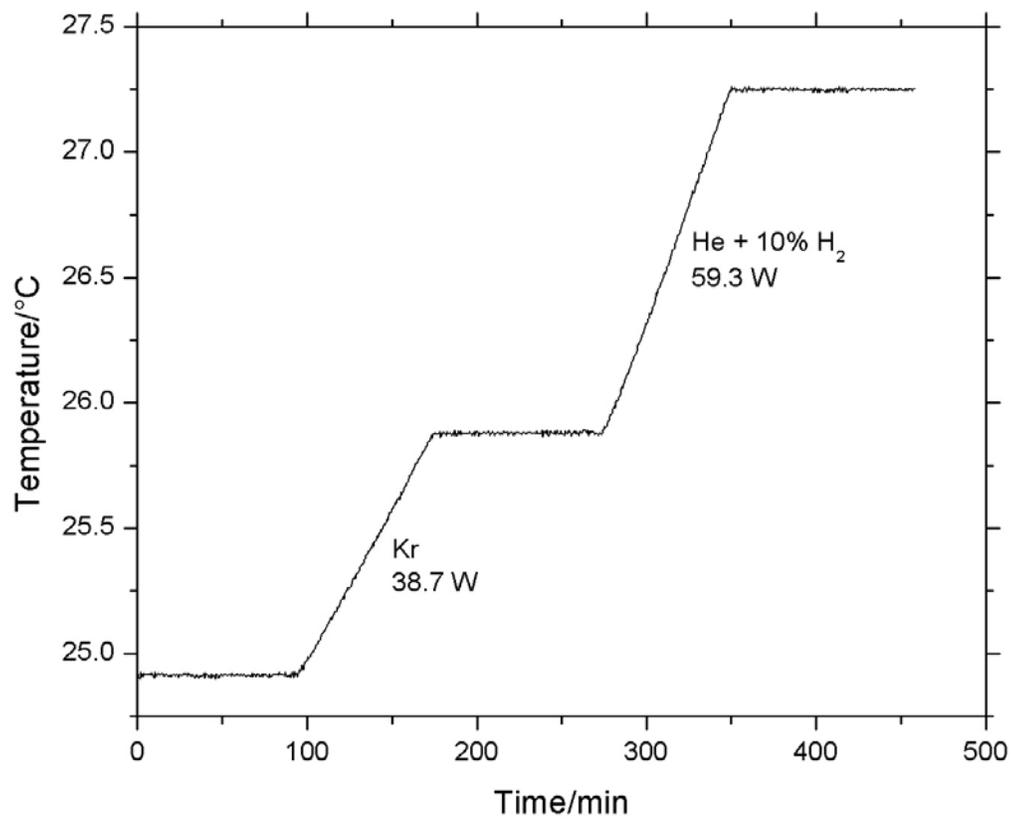

Fig. 4

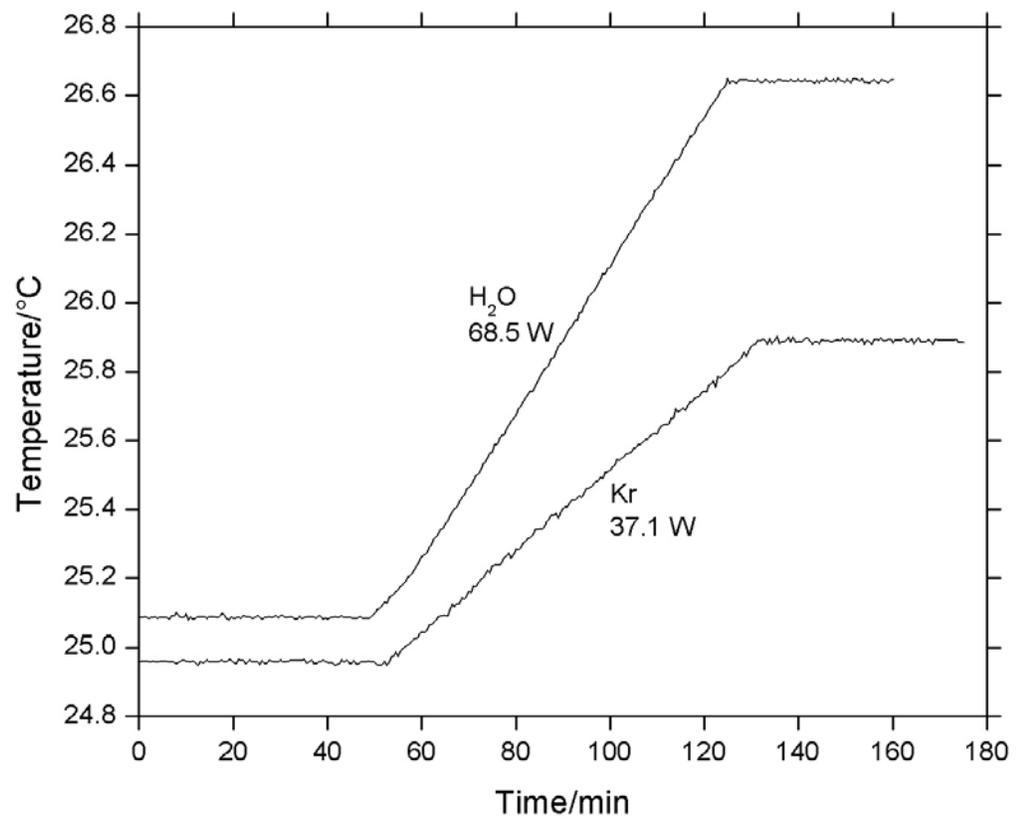

Fig. 5